\documentclass[preprint]{acmart}

\usepackage{booktabs} 

\begin{document}
\title{Applications of Machine Learning in Cryptography: A Survey}

\author{Mohammed M. Alani}
\orcid{0000-0002-4324-1774}
\affiliation{%
  \institution{Khawarizmi International College}
  \city{Abu Dhabi}
  \country{United Arab Emirates}
}
\email{m@alani.me}

\copyrightyear{2019} 
\acmYear{2019} 
\setcopyright{acmlicensed}
\acmConference[ICCSP 2019]{2019 the 3rd International Conference on Cryptography, Security and Privacy}{January 19--21, 2019}{Kuala Lumpur, Malaysia}
\acmBooktitle{2019 the 3rd International Conference on Cryptography, Security and Privacy (ICCSP 2019), January 19--21, 2019, Kuala Lumpur, Malaysia}
\acmPrice{15.00}
\acmDOI{10.1145/3309074.3309092}
\acmISBN{978-1-4503-6618-2/19/01}

\renewcommand{\shortauthors}{M.M. Alani}

\begin{abstract}
Machine learning techniques have had a long list of applications in recent years. However, the use of machine learning in information and network security is not new. Machine learning and cryptography have many things in common. The most apparent is the processing of large amounts of data and large search spaces.\\
In its varying techniques, machine learning has been an interesting field of study with massive potential for application. In the past three decades, machine learning techniques, whether supervised or unsupervised, have been applied in cryptographic algorithms, cryptanalysis, steganography, among other data-security-related applications.\\
This paper presents an updated survey of applications of machine learning techniques in cryptography and cryptanalysis. The paper summarizes the research done in these areas and provides suggestions for future directions in research.
\end{abstract}

%
%

\keywords{cryptography, cryptanalysis, machine learning}

\maketitle
\section{Introduction}
In 1991, Ronald Rivest, one of the creators of RSA, presented an invited talk in ASIACYPT'1991 about cryptography and machine learning\cite{ref3}. In his talk, Rivest discussed the similarities and differences between machine learning and cryptography, and how each area can impact the other. Since Ronald Rivest spoke about the “cross-fertilization” of the fields of machine learning and cryptography, the area has gained massive attention.\\
In general, machine learning and cryptanalysis have more in common that machine learning and cryptography. This is due to that they share a common target; searching in large search spaces. A cryptanalyst's target is to find the right key for decryption, while machine learning's target is to find a suitable solution in a large space of possible solutions.\\
The applications of different machine learning techniques have gained growing attentions over the past years. The areas of research for these applications were mainly in the directions list below:
\begin{enumerate}
\item Applications in cryptography
\begin{enumerate}
\item Cryptosystems based on machine learning
\item Classification of encrypted traffic
\end{enumerate}
\item Applications in cryptanalysis
\begin{enumerate}
\item Cryptanalysis of encryption algorithms
\item Attacks based on machine learning
\end{enumerate}
\end{enumerate} 
Other areas of research do exist, like privacy preservation\cite{ref16, ref17,ref20,ref30, ref41, ref42}, quantum machine learning\cite{ref33, ref43}, and the manipulation of machine learning to mislead its application to wrong classification.\\
Mutual research of machine learning and cryptography is not a new area. That is why you will find in the survey presented here publications that go back to 1990s. In 1990, Kearns discussed in his dissertation the crpytographical aspects of machine learning and how computational complexity can be calculated\cite{ref21}. This publication paved the way for future research that involved cryptography and machine learning.\\
In addition to cryptography and cryptanalysis, machine learning has a wide range of applications in relation to information and network security. A none-exhaustive list of examples found here:
\begin{enumerate}
\item Using machine learning to develop Intrusion Detection System (IDS) \cite{ref23, ref32, ref24}
\item Botnet detection\cite{ref25}
\item Network anomaly detection\cite{ref28}
\item Malware analysis and detection\cite{ref34, ref38}
\item Homomorphic encrpytion applications\cite{ref35}
\item Attacks on Physical Unclonable Functions (PUFs)\cite{ref36}
\item Malicious code classification and identification\cite{ref40, ref39}
\end{enumerate}

\section{Machine Learning and Its Applications}
Machine learning happens when we need a machine(computer) to learn to solve a problem based on, usually large amounts of, data previously fed into the machine. Machine learning can be a good solution finder for problems for which we do not have a clear algorithm to solve. For example, when we want the computer to be able to detect spam emails, there is not clear algorithm that is 100\% accurate in finding spam. Hence, machine learning can be a closer-to-optimum solution when we feed the machine hundreds or thousands of spam and non-spam examples. Gradually, the machine will learn to me more and more accurate in detecting spam emails. The larger the data used for learning becomes, the more accurate the classification becomes\cite{ref1}.\\
With around 2.5 quantillion bytes of data being generated every day\cite{ref2}, legacy database systems become incapable of processing such amounts of data. Unless new technologies are used to transfer, store, and process such large amounts of data, all of the data being collected will be unusable.\\
The consumption of this data in a useful commercial application requires data processing techniques that are design specifically to handle these large amounts of data. Whether the applications are commercial, like processing user access patterns to show highly-targeted ads, or for health uses, like predicting illnesses or illness prediction patterns, or even political, like predicting civil unrest, machine learning can be of great help in producing useful information from extremely large amounts of data.\\ 
Classification is one of the most widely used applications of machine learning\cite{ref1}. A common example of classification is the classification that banks use for loans; low-risk, and high-risk. A bank needs to be able to properly classify a loan into one of the two categories. Generally, high-risk loans are loans that have a high probability of defaulting payments. On the other hand, low-risk loans are loans that are highly expected to be paid back on time, and in regular fashion. The data fed into the classification software, which is basically historical data about the customer, like credit score, is used as the input. This input will be used in training the software to produce a rule that is relevant to several factors within the situation of the loan applicant. This rule will help the bank decide whether this loan is a high-risk or a low-risk loan.\\
Another example of machine learning applications is regression. Regression is the type of problem that produces a number based on multiple inputs. An example is when a government would like to predict the global prices of oil. The input to this system is all elements that are expected to impact the oil prices, like global production rates, current price, last several month's prices, season (winter, spring,..etc). The output would be a specific number driven from the inputs. However, there has to be training that would enable the system to be more accurate gradually and to learn the impact of change in each of the input elements.\\
Learning associations is also one of the applications of machine learning. For example, analysis of shopping baskets data can produce useful information to supermarkets that they can use to improve  their sales. When the supermarket system analyzes the data of thousands of shopping baskets, it will be able to produce useful association information. The system would help the management know that 80\% of people who buy soda, buy potato chips as well. Hence, placing them in the same aisle can be a useful action to help boost items sales. The system would be able to recognize that most people buying canned food are likely to buy disposable plates. Hence, packaging them together or placing them in the same isle can increase their sales.\\
Unsupervised learning can also be used in machine learning. When there is no reference output to compare to, the learning is done through input data only. This type of learning or training is called unsupervised. A common type of unsupervised learning is called clustering. Clustering is used in finding structures or clusters in input data. Companies might use this method to process past data of customers to try to find density profiles for customers.\\
Reinforcement learning can also be used in machine learning applications. In certain applications, the output of the system is a sequence of actions. A signle action by itself is not important, but the sequence of the correct action. For example, when playing a game, a single move might not be important but a sequence of correct actions can lead to a successful accomplishment.\\

\section{Applications in Cryptography}
\label{apps-in-crpyto}
Rosen-Zvi et. al. proposed, in 2002, that mutual learning in a tree parity machine can be used as public-key cryptosystem\cite{ref22}. The paper explains how the tree parity machine has a potential in being used as a public-key cryptosystem by using its synchronization state as the key in a certain encryption and decryption rule. This "key" can be exchanged in public without the need for prior communication. The receiving side only needs to perform a finite number if steps of input-output exchange to achieve convergence to a synchronization state.\\
In 2005, Klien et. al. introduced a new phenomenon named mutual learning\cite{ref45}. This mutual learning phenomenon can be used in cryptography. Mutual learning can help the two sides of communication to create a common secret key over a public channel. The two parties have advatage over the attacker that the attacker will be learning one-way, which would make it virtually impossible to create the same secret key that the two sides would have.\\
Blum, in 2007, discussed the possibilities of connecting machine learning and cryptography\cite{ref29}. The paper states that at a technical level, strong connections between techniques of machine learning and techniques used in cryptography exist. As an example, the author discussed "Boosting" technique, which is a machine learning method designed to extract as much power as possible out of a learning algorithm. This can be connected to methods for amplifying cryptosystems.\\
In 2009, Al-Shammari and Zincir-Heywood introduced a classification technique, based on machine learning, to classify encrypted traffic\cite{ref8}. The work was done to assess the robustness of machine learning classification of encrypted traffic. the work focused on flow-based without employing commonly used features like Internet Protocol(IP) addresses, port numbers, and payload information. The results of this study have shown that C4.5 learning algorithm outperformed other algorithms like RIPPER, Naive Bayesian, support vector machine, and AdaBoost.\\
In 2011, Huan, et. al, presented an invited talk about the adversarial machine learning\cite{ref14}. The paper introduced taxonomy for classifying attacks against online machine learning algorithms. In addition, the paper explored vulnerabilities in machine learning algorithms and their counter measures. The paper also presented two models by which adversaries capabilities can be modeled. The paper discussed thoroughly a specific type of attacks name exploratory integrity attacks.  In this type of attacks, the adversary attempts to passively circumvent the learning mechanism to exploit blind spots in the learner that allow miscreant activities to go undetected. This type of attacks can be used in different applications of machine learning.\\
In 2012, Schaathun presented a book with a collection of ideas to use machine learning in steganalysis\cite{ref4}. The book describes ways in which machine learning can be employed in countering steganography. The main idea presented was to classify objects into steganograms or clean documents using machine learning classification capabilities.\\
Graepel, et. al. presented, in 2012, a system by which it was possible to delegate the execution of the machine learning algorithm to a computing service while maintaining confidentiality through the use of a leveled homomorphic encryption scheme\cite{ref13}. The rationale behind this approach was to employ higher performance computing services, like cloud computing, to improve the speed of machine learning and the capability to process larger amounts of data.\\
Sagar and Kumar introduced, in 2014, a symmetric key encryption algorithm that was based on counter propagation network\cite{ref10}. The algorithm proposed is a symmetric block cipher where data is converted into bits and passed through the neural network as the plainttext, and the resulting output of the unsupervised learning process is considered the output. Although the principle is not presented clearly in the paper, the direction can be further explored.\\
Bost, et al. presented three major classification protocols that rely on machine learning in the classification of encrypted data, in 2015\cite{ref7}. The three protocols presented, hyperplane decision, Naïve Bayesian, and decision trees, were combined with AdaBoost. The work produced privacy-preserving classifiers. The proposed classifiers were tested on medical datasets and have proven to achieve proper classification with satisfactory efficiency.\\

\section{Applications in Cryptanalysis}
Machine learning techniques were also applied to perform side-channel attacks. In 2011, Hospodar, et. al, presented a first study that suggested utilizing machine learning in side-channel attacks\cite{ref12}. The proposed system used Least Squares Support Vector Machine (LS-SVM) learning algorithm with the side-channel being the power consumption, wuth a target of software implementation of Advanced Encrpytion Standard (AES). The study have shown that the choice of parameters of the machine learning algorithm strongly impacts the results.\\
Another side-channel and machine learning coupling was published in 2014. Lerman, Bontempi, and Markowitch proposed the use of machine learning techniques to improve accuracy of side-channel attacks\cite{ref6}. As side-channel attacks rely on the physical measurements of hardware implementations of cryptosystems, there are always certain parametric assumptions that these attacks rely on. The proposed usage of machine learning relaxes such assumptions and deal with high dimensional feature vectors.\\
In 2012, Alani introduced a known-plaintext attack that employed a neural network to perform cryptanalysis\cite{ref9}. The proposed attack trains a neural network to decrypt ciphertext without knowing the encryption key. The attack was successful in reducing the time and the known plaintext-ciphertext pairs needed for Data Encryption Standard (DES) and Triple-DES to a great extent compared to other known-plaintext attacks.\\
In continuation of the work previously presented by Alani in \cite{ref9}, Jayachandiran implemented a similar attack on a lightweight cipher named Simon\cite{ref11}. However, the attack this time was aimed at finding the key rather than the plaintext. The proposed neural network was tested on reduced-round version of the cipher, and on the full-round version as well. Network configurations were also changed to try to find achieve the highest possible accuracy.\\
Conti, et al., in 2015, published research on analysis of Android encrypted network traffic\cite{ref5}. This analysis was focused on identifying user action, in spite of being encrypted. In this research, the adversary does not interact actively with the target, but eavesdrop on encrypted traffic. The collected encrypted traffic is then analyzed using advanced machine learning techniques to figure out the actions performed by the user. The proposed system achieved accuracy of up to 95\% in identifying the user actions.\\
In 2016, Maghrebi et. al. published their research on the use of deep learning in side-channel attacks\cite{ref31}. The paper examines the possibility of a more sophisticated profiling techniques to reduce assumptions in template attacks. In this paper, deeplearning techniques were used to produce more accurate results in side-channel attacks on AES. The results stated in the paper confirmed the advatages of this technique implemented on protected and unprotected implementations of AES.\\
Yu et. al. presented, in 2016, a technique that can prevent machine learning from being used as an attack tool in PUFs for lightweight authentication\cite{ref44}. The paper introduced a lightweight PUF-based authentication approach that is couple with a lockdown technique to prevent machine learning from successfully obtaining the new challenge-response pair.\\  

\section{Attacks on Machine Learning}
Barreno et. al. published, in 2006, a paper discussing whether machine learning can be secure or not\cite{ref26}. The paper introduced a taxonomy of different types of attacks on machine learning techniques and systems. The paper also introduced defenses against those attacks with an analytical model showing attacker's work function.\\
In 2010, Barreno et. al. also published a detailed paper building on their previous work\cite{ref27}. In this paper, the authors expanded their taxonomy of attacks and shown how the classes identified influence the cost for the attacker and defender. The paper also presented a thorough survey of attacks on machine learning systems and techniques discussed in previous literature. It illustrates the taxonomy by showing how it can guide attacks against SpamBayes, the statistical spam filter.\\
Biggio, et. al discussed, in 2013, a type of attack named evasion attack\cite{ref15}. Although to some extent, similar to exploratory integrity attacks mentioned in section \ref{apps-in-crpyto}, this attack relies on injecting adversarial data in the training data used in a machine-learning-based system. The paper states that it is important for machine learning to have thorough vetting of its resistance to adversarial data.\\
In 2013 as well, Ateniese et. al. introduced a method by which machine learning classifiers can be exploited to get information by attackers\cite{ref19}. The paper focused in statistical information that can be unconsciously or maliciously revealed from machine learning classifiers. In this paper, a novel meta-classifier was built and trained to hack other classifiers and obtain meaningful information about their training sets. This kind of attack can be used to build a more effective classifier or to obtain trade secrets from competitors by violating intellectual property rights.\\
However, learning approaches can be evaded by adversaries, who change their behavior in response to the learning methods. To-date, there has been limited research into learning techniques that are resilient to attacks with provable robustness guarantees The Perspectives Workshop, "Machine Learning Methods for Computer Security" was convened to bring together interested researchers from both the computer security and machine learning communities to discuss techniques, challenges, and future research directions for secure learning and learning-based security applications. As a result of the twenty-two invited presentations, workgroup sessions and informal discussion, several priority areas of research were identified. The open problems identified in the field ranged from traditional applications of machine learning in security, such as attack detection and analysis of malicious software, to methodological issues related to secure learning, especially the development of new formal approaches with provable security guarantees. Finally a number of other potential applications were pinpointed outside of the traditional scope of computer security in which security issues may also arise in connection with data-driven methods. Examples of such applications are social media spam, plagiarism detection, authorship identification, copyright enforcement, computer vision (particularly in the context of biometrics), and sentiment analysis.

Peprnot, et. al. presented, in 2016, a thorough study of security and privacy in machine learning\cite{ref18}. The paper introduces a comprehensive threat model for machine learning along with proper categorization of attacks and defenses within an adversarial framework. The adversarial settings of training were categorized to two categories; targeting privacy, and targeting integrity. On the other hand, inferring in adversarial settings was categorized in to two categories as well; white-box adversaries, and black-box adversaries. The paper also discussed how a robust, private, and accountable machine learning model can be achieved.\\

\section{Conclusion and Future Directions}
In this paper, we presented a survey of applications of machine learning in cryptography and cryptanalysis. The review of the literature indicated that there is a huge room for improvement in these research areas. The paper have also discussed security attacks on machine learning techniques and machine-learning-based systems.\\

Future directions in machine learning and cryptography may include:
\begin{enumerate}
\item Employing machine learning algorithms in symmetric and asymmetric cryptosystems design. Perhaps with the advancements in Artificial Intelligence (AI) we can witness two AI-based systems design their own cryptosystem.
\item Employing machine learning in privacy-reserving training based on obscured data sets.
\item The use of machine learning techniques in cryptanalysis to extract decryption keys from ciphertext blocks.
\item Connecting machine learning with existing cryptanalysis techniques (like differential or linear cryptanalysis) to improve their efficiency in finding solutions in the search space. 
\end{enumerate}

\bibliographystyle{ieeetr}
\bibliography{references}
\end{document}